# Profiles for ion activities are multiphasic, not curvilinear

Per Nissen

Norwegian University of Life Sciences
Department of Ecology and Natural Resource Management
P. O. Box 5003, NO-1432 Ås, Norway

Bob Eisenberg

Department of Physiology and Biophysics
Rush University
Chicago, USA

2018

# Abstract

In agreement with previous findings (Nissen 2017a,b), data (Wilczek-Vera et al. 2004) for the ion activities of a variety of simple inorganic compounds are shown to be precisely represented as multiphasic, i.e. as a series of straight lines separated by discontinuous transitions.

# Introduction

A wide variety of processes and phenomena, non-biological as well as biological, has been conventionally, and maybe wrongly, represented by curvilinear profiles. Thus, ion uptake in plants has been found to be multiphasic when plotted in linear transformations of the Michaelis-Menten equation (Nissen 1971, 1974, 1991, 1996), as have published data for other transport, binding and enzyme systems (Nissen and Martín-Nieto 1998). Such profiles have now been found also for activation of ion channels, binding, pH, folding/unfolding, effects of various interactions and of chain length (Nissen 2015a,b, Nissen 2016a-d). These processes have little in common beyond the ions involved. The finding (Nissen 2017a,b) of multiphasic profiles for ion activities of NaCl, KCl, NaBr and KBr at 15, 25, 35 and 45°C (data of Lee et al. 2002) indicate that they are somehow also causing the multiphasic profiles in more complex systems. In the present paper, data (Wilczek-Vera et al. 2004) for cation and anion activation are similar for some compounds, but the multiphasic profiles can be widely different for other compounds.

# Analysis and Results

The data for $m$ and $\gamma$ in Tables 1-7 in Wilczek-Vera et al. (2004) have been plotted over the entire concentration range both for cation and anion activities. For KOH, plots at low concentrations are also shown. The findings (number of phases, concentrations at or over which the transitions occur, occurrence of jumps and parallel lines, and any special comments) are given briefly under each pair of plots. The r values for the straight lines are given on the plots, as are the slopes ± SE (or only slopes). The figures group for the most part quite logically together on the same page.

A major finding in the analysis of these and many other data sets giving multiphasic profiles is the extreme precision of the data. For the present data there is a total of 130 lines with 3 or more points. Of these, 93% have absolute r values of 0.995 or higher, 75% have values of 0.999 or higher, and 30% have values of 0.9999 or higher. The few low values are for the most part due to
the lines being horizontal or almost so.

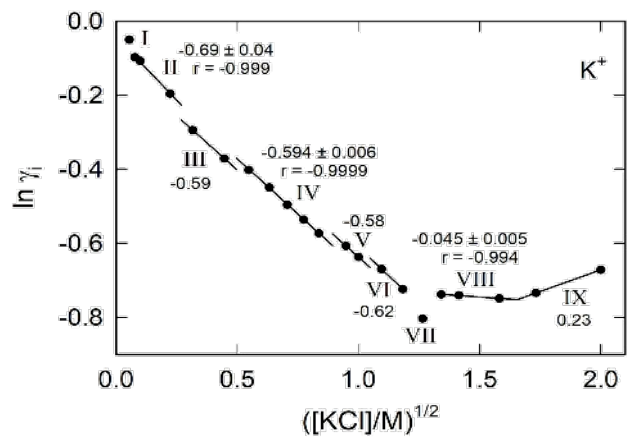 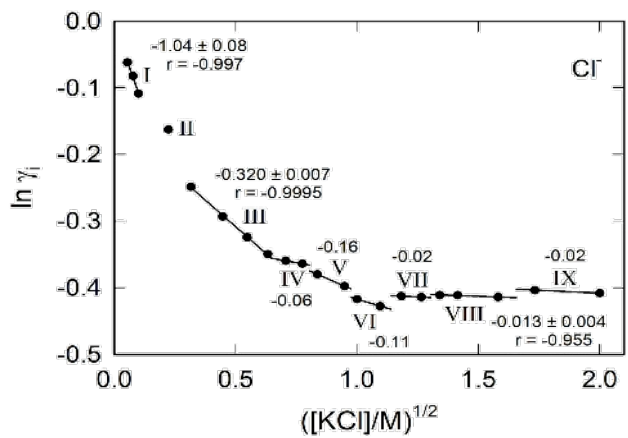

Figs 1A and 1B. Data from Table 1.
$K^+$: Nine phases. Transitions: 0.05-0.08, 0.22-0.32 (jump), 0.45-0.55 (jump), 0.84-0.95 (jump), 1.00-1.10 (jump), 1.18-1.26, 1.26-1.34, 1.66. Lines III-VI and possibly also line II are parallel.
$Cl^-$: Nine phases. Transitions: 0.10-0.22, 0.22-0.32, 0.65, 0.77-0.84 (jump), 0.95-1.00 (jump), 1.10-1.18 (jump), 1.26-1.34 (tiny jump), 1.58-1.73 (jump). Lines V and VI may be parallel, lines VII-IX may be a single horizontal line. Single line for lines IV-VI: r = -0.989.

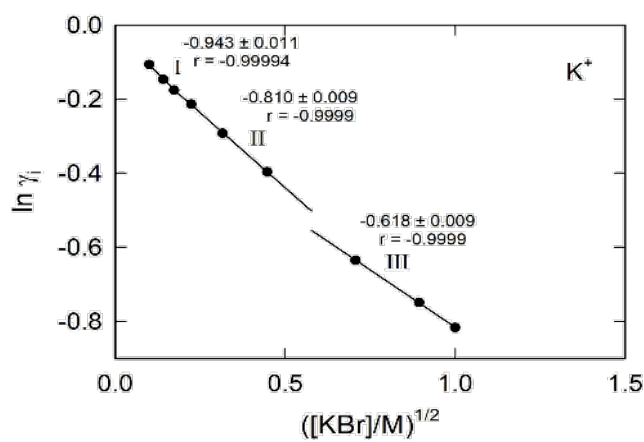 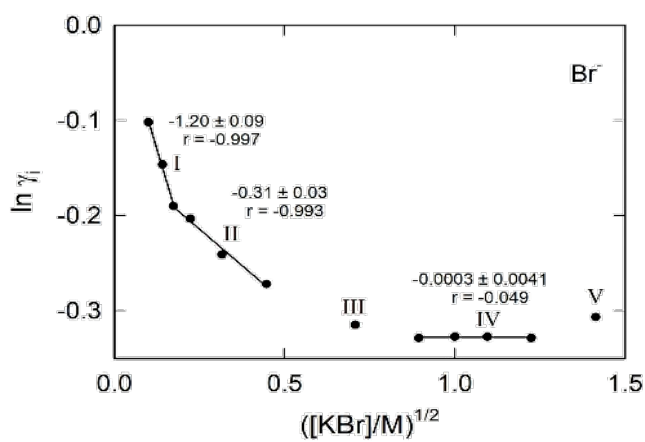

Figs 2A and 2B. Data from Table 1.
$K^+$: Three phases (up to 1.00). Transitions: 0.17, 0.45-0.71 (jump). Very straight lines.
$Br^-$: Five phases (up to 1.41). Transitions: 0.17, 0.45-0.71, 0.71-0.89, 1.22-1.41. Line IV is horizontal. Phases III and V consist of only a single point.

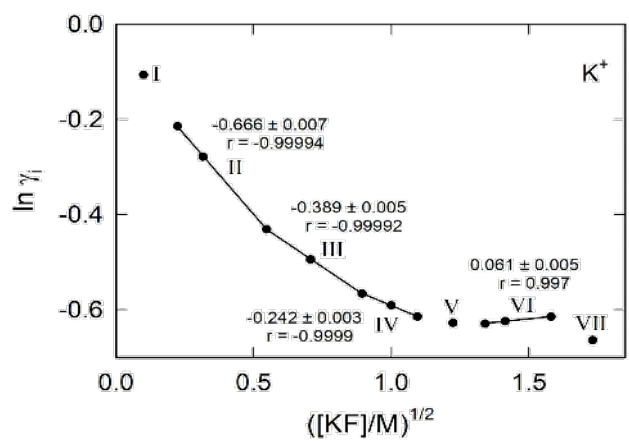 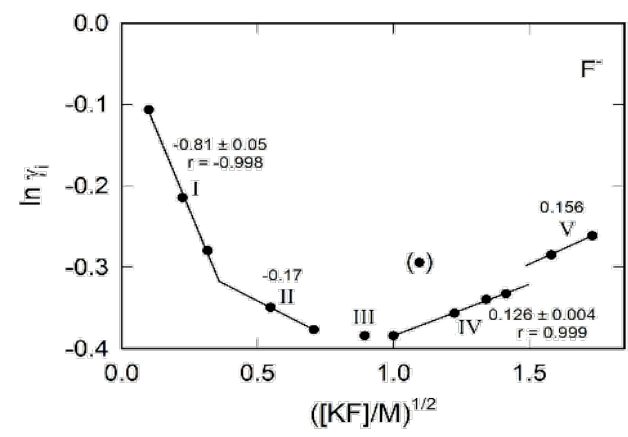

Figs 3A and 3B. Data from Table 1.
$K^+$: Seven phases. Transitions: 0.10-0.22, 0.55, 0.89, 1.10-1.22, 1.22-1.34, 1.58-1.73. Phases I, V and VII consist of only a single point.
$F^-$: Five phases. Transitions: 0.36, 0.71-0.89, 0.89-1.00, 1.41-1.58 (marked jump). The point at 1.10 is probably in error. Lines IV and V are about parallel.

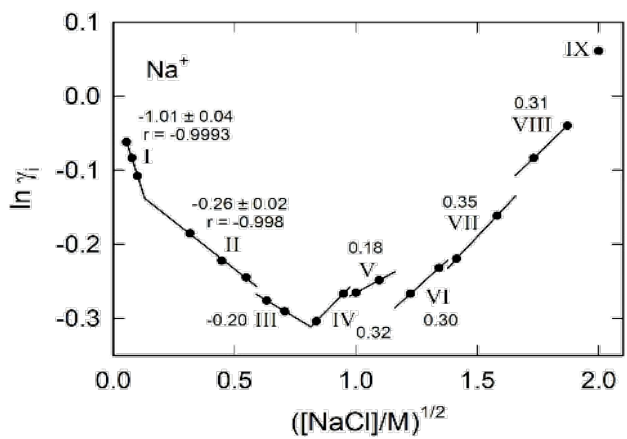 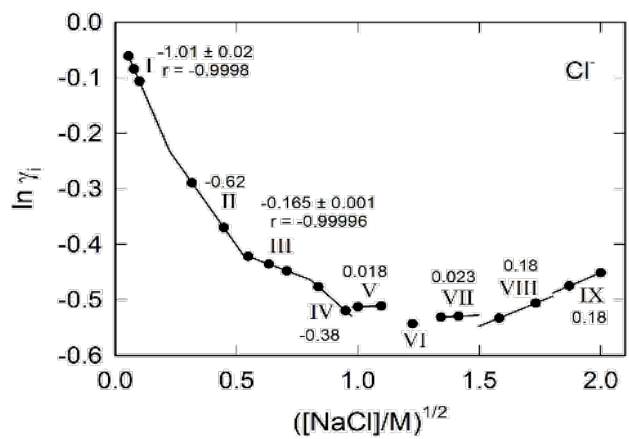

Figs 4A and 4B. Data from Table 2.
Na$^+$: Nine phases. Transitions: 0.13, 0.55-0.63 (jump), 0.81, 0.95-1.00 (jump), 1.10-1.22 (large jump), 1.34-1.41 (jump), 1.58-1.73 (large jump), 1.87-2.00. Lines II and III are about parallel, as are lines IV, VI, VII and VIII.
Cl$^-$: Nine phases. Transitions: 0.22, 0.53, 0.80, 0.95-1.00 (jump), 1.10-1.22, 1.22-1.34, 1.41-1.58 (jump), 1.73-1.87 (jump). Lines V and VII are parallel, as are lines VIII and IX.

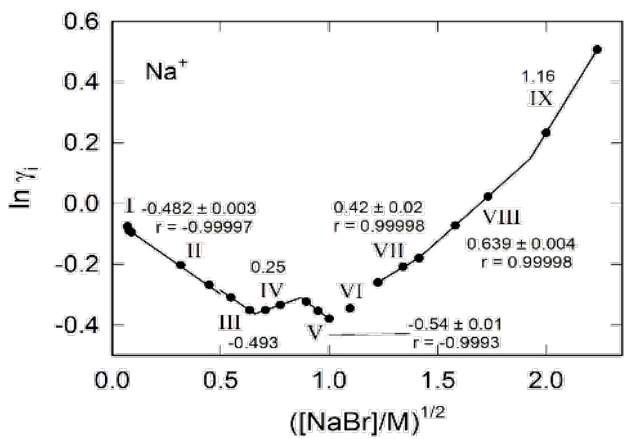 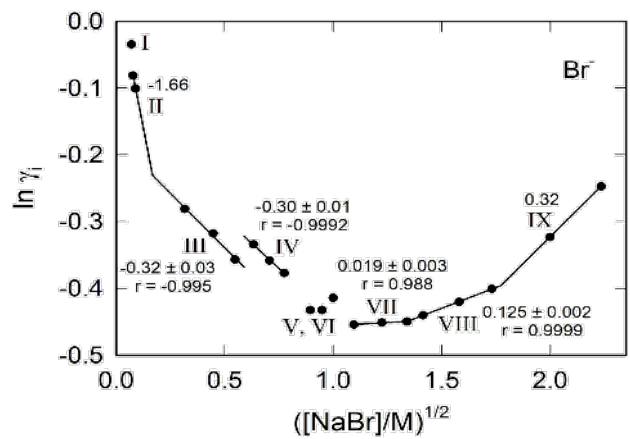

Figs 5A and 5B. Data from Table 2.
Na$^+$: Nine phases. Transitions: 0.07-0.08, 0.45-0.55 (jump), 0.66, 0.87, 1.00-1.10, 1.10-1.22, 1.41, 1.93. Lines II and III are parallel. A single line for phases I and III: r = -1.000000.
Br$^-$: Nine phases. Transitions: 0.07-0.08, 0.17, 0.55-0.63 (large jump), 0.77-1.10 (phases V and VI), 1.34, 1.77. Lines III and IV are parallel.

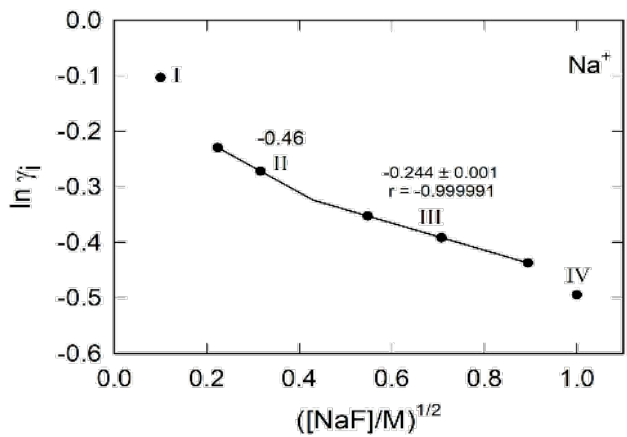 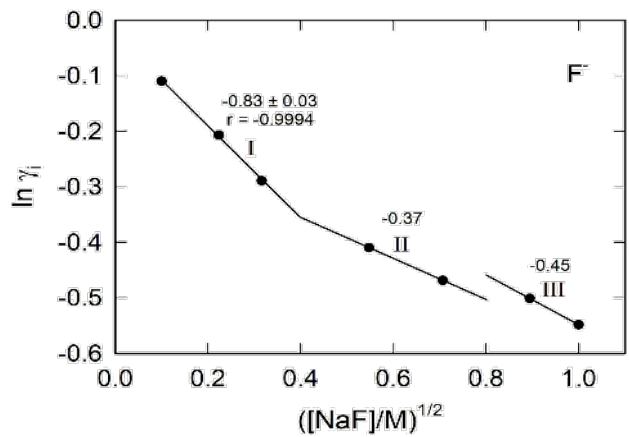

Figs 6A and 6B. Data from Table 2.
Na$^+$: Four phases. Transitions: 0.10-0.22, 0.43, 0.89-1.00.
F$^-$: Three phases. Transitions: 0.40, 0.71-0.89 (large jump).

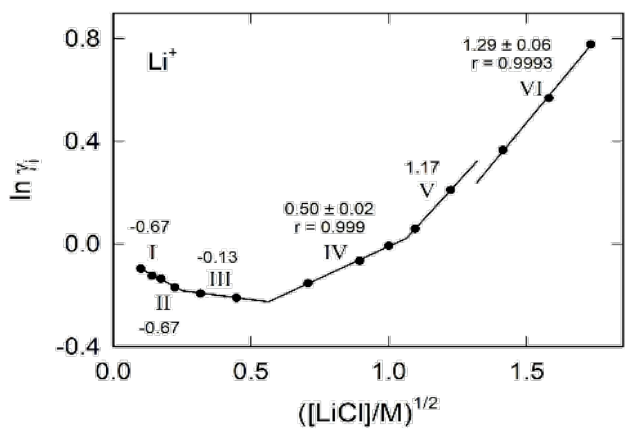
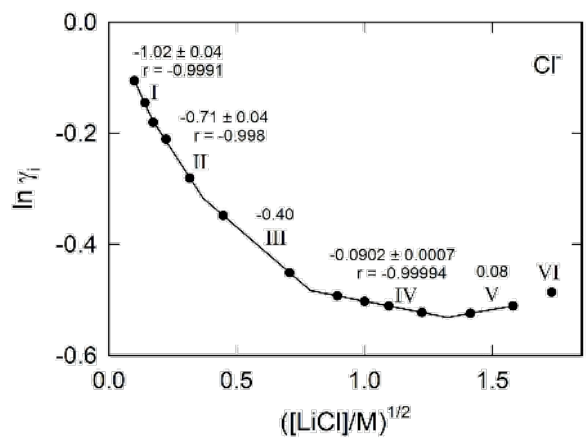

Figs 7A and 7B. Data from Table 3.
Li$^+$: Six phases. Transitions: 0.14-0.17 (tiny jump), 0.25, 0.56, 1.06, 1.22-1.41 (jump). Lines I and II are parallel, lines V and VI approximately so.
Cl$^-$: Six phases. Transitions: 0.17, 0.37, 0.79, 1.32, 1.58-1.73.

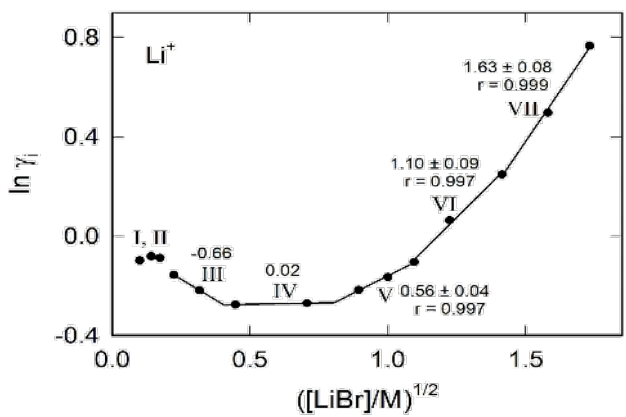
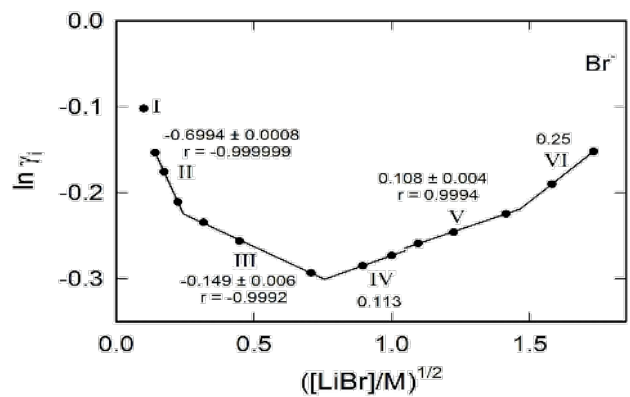

Figs 8A and 8B. Data from Table 3.
Li$^+$: Seven phases (but I and II cannot be assigned). Resolvable transitions: 0.43, 0.81, 1.10, 1.41.
Br$^-$: Six phases. Transitions: 0.10-0.14, 0.24, 0.76, 1.00-1.10 (tiny jump), 1.47. Lines IV and V are parallel.

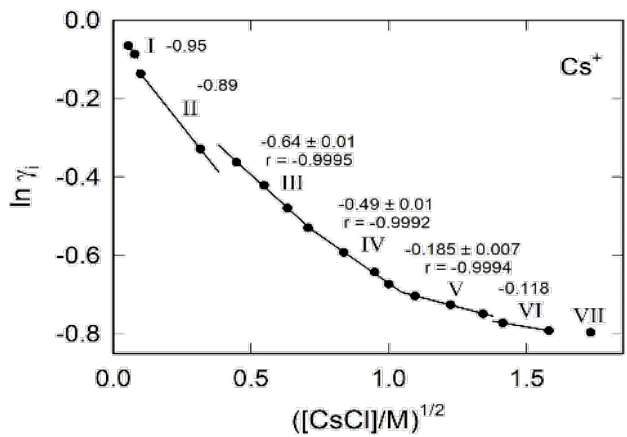
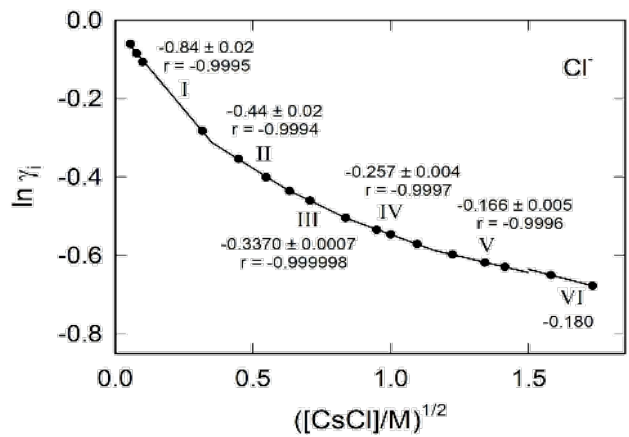

Figs 9A and 9B. Data from Table 3.
Cs$^+$: Seven phases. Transitions: 0.08-0.10 (tiny jump), 0.32-0.45 (jump), 0.71, 1.06, 1.34-1.41 (tiny jump), 1.58-1.73. Lines I and II are about parallel, as are, roughly, also lines V and VI.
Cl$^-$: Six phases. Transitions: 0.35, 0.63, 0.84, 1.15, 1.41-1.58 (tiny jump). Lines V and VI are parallel.

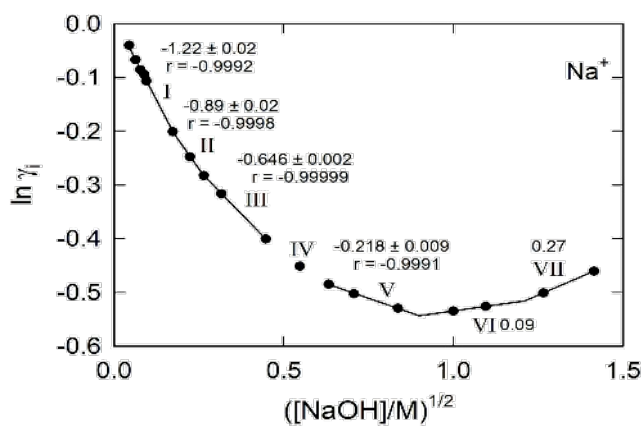
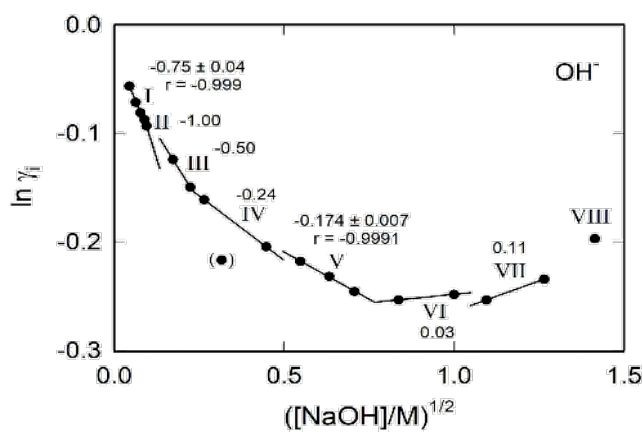

Figs 10A and 10B. Data from Table 4.
$Na^+$: Seven phases. Transitions: 0.17, 0.26, 0.45-0.55, 0.55-0.63, 0.90, 1.21.
$OH^-$: Eight (or seven) phases. Transitions: 0.08, 0.09-0.17 (jump), 0.24, 0.45-0.55 (jump), 0.78, 1.00-1.10 (jump), 1.26-1.41. A single line for phases I and II will have r = -0.998. The value for point 9 is probably in error.

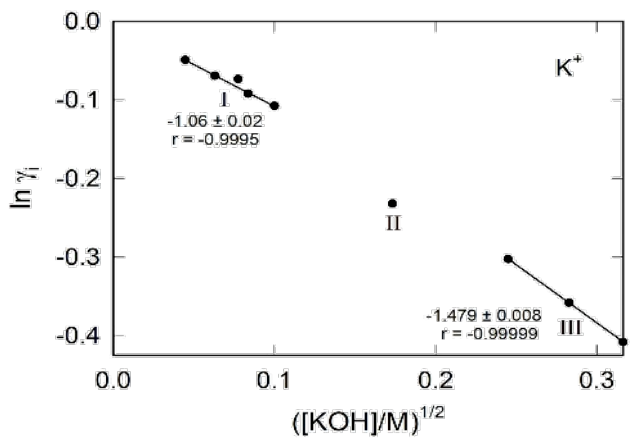
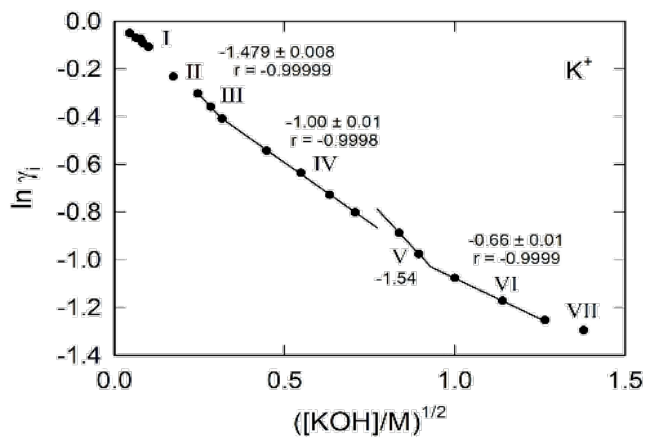

Figs 11A and 11B. Data from Table 4.
$K^+$: Seven phases (overlapping plots. Transitions: 0.10-0.17, 0.17-0.24, 0.32, 0.71-0.84 (jump), 0.93, 1.26-1.38. Point 3 may be in error and has not been included in the calculation of phase I. (Its inclusion will give r = -0.976.) All phases form roughly a straight line.

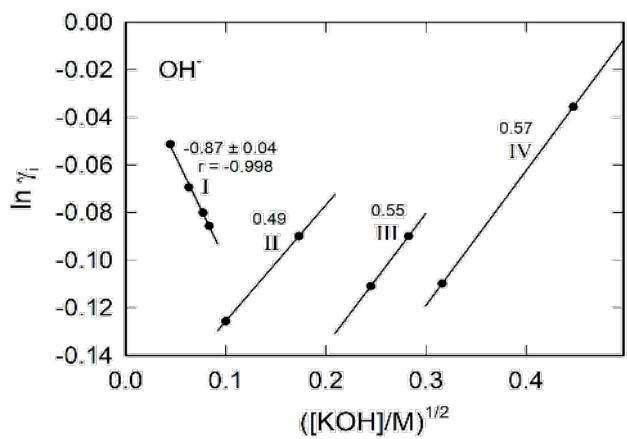
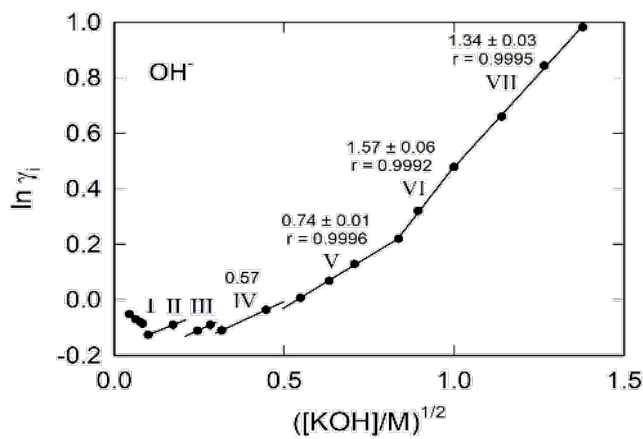

Figs 11C and 11D. Data from Table 4.
$OH^-$: Seven phases (overlapping plots): Transitions: 0.08-0.10 (jump), 0.17-0.24 (jump), 0.28-0.32 (jump), 0.45-0.55 (tiny jump), 0.84, 1.00. Lines II-IV are parallel or closely so.

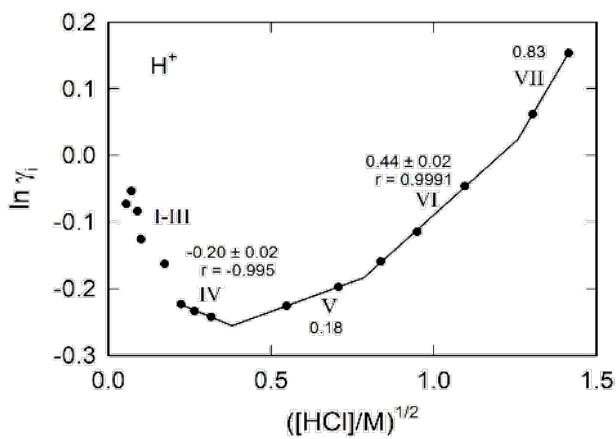
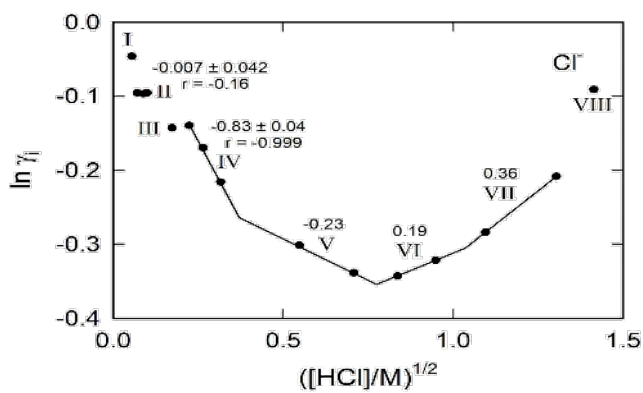

Figs 12 A and 12B. Data from Table 4.
H$^+$: Probably three phases in the range of the first 5 points, with the remaining transitions: 0.17-0.22, 0.38, 0.78, 1.26.
Cl$^-$: Probably also three initial phases for a total of eight phases, with the remaining transitions: 0.17-0.22, 0.37, 0.77, 1.04, 1.30-1.41. Line II is horizontal.

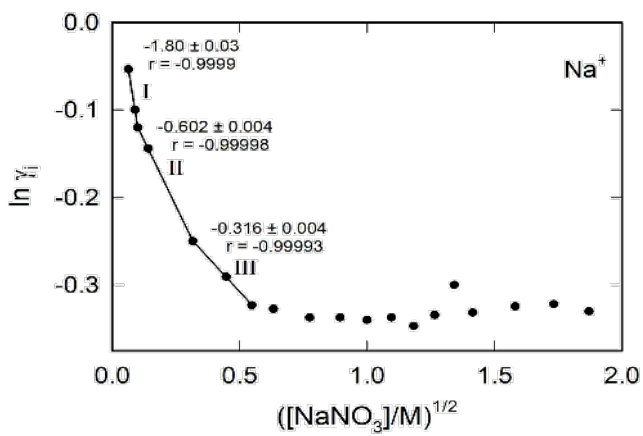
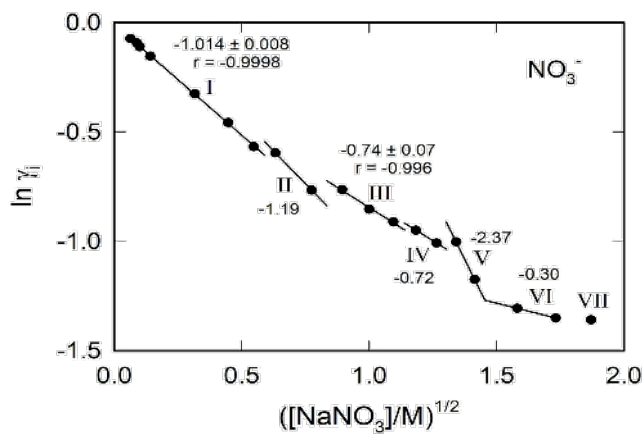

Figs 13A and 13B. Data from Table 5.
Na$^+$: Only 3 phases can be reliably resolved. Transitions: 0.10, 0.32.
NO$_3^-$: Seven phases. Transitions: 0.55-0.63 (tiny jump), 0.77-0.89 (jump), 1.10-1.18 (tiny jump), 1.26-1.34 (jump), 1.45, 1.73-1.87. Lines III and IV are parallel, lines I and II approximately so.

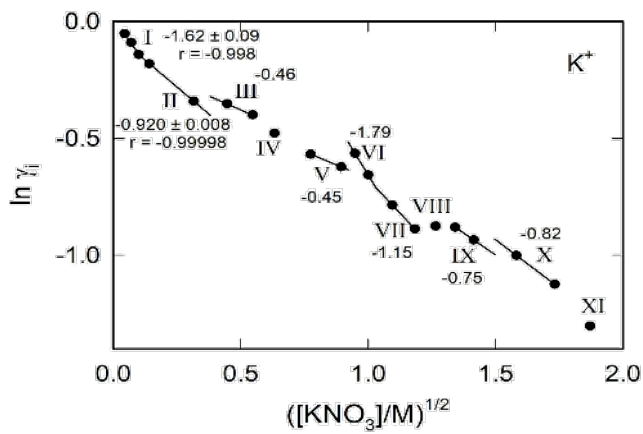
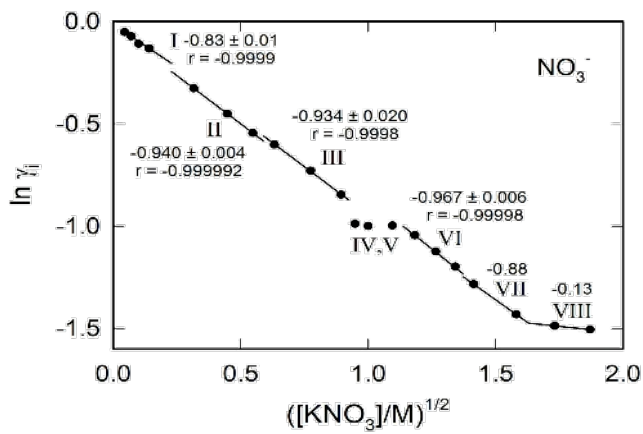

Figs 14A and 14B. Data from Table 5.
K$^+$: Eleven phases. Transitions: 0.10, 0.32-0.45 (jump), 0.55-0.63, 0.63-0.77, 0.89-0.95, 1.03, 1.18-1.26, 1.26-1.34, 1.41-1.73, 1.73-1.87. Lines III and V are parallel, as are approximately also lines IX and X.
NO3-: Eight phases. Transitions: 0.14-0.32 (jump), 0.55-0.63 (tiny jump), 0.89-1.18 (phases IV and V), 1.14, 1.62. Point 3 is

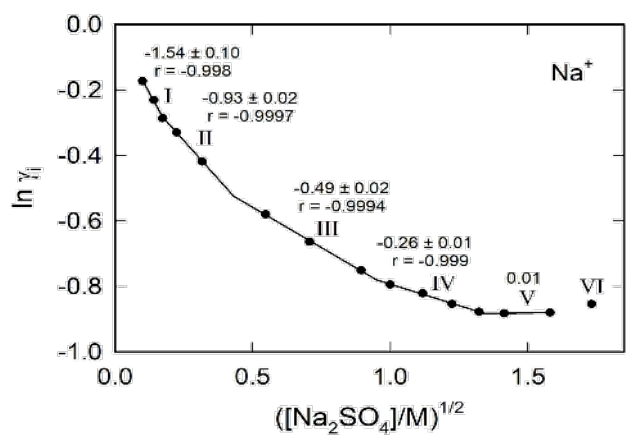
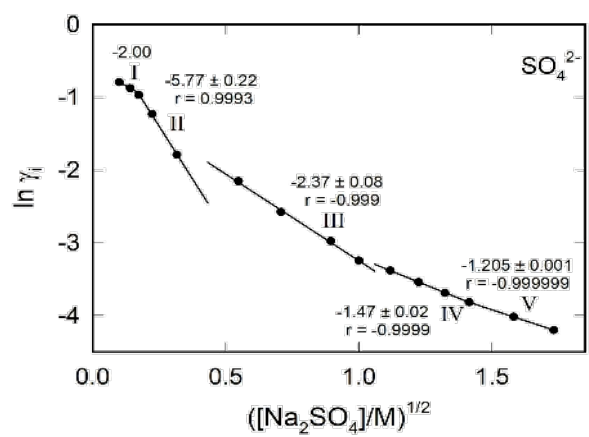

Figs 15A and 15B. Data from Table 5.
$Na^+$: Six phases. Transitions: 0.17, 0.43, 0.95, 1.34, 1.58-1.73.
$SO_4^{2-}$: Five phases. Transitions: 0.17, 0.32-0.55 (large jump), 1.00-1.12 (jump), 1.41.

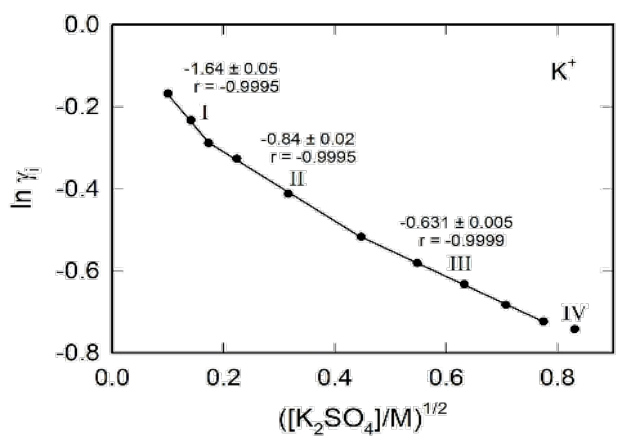
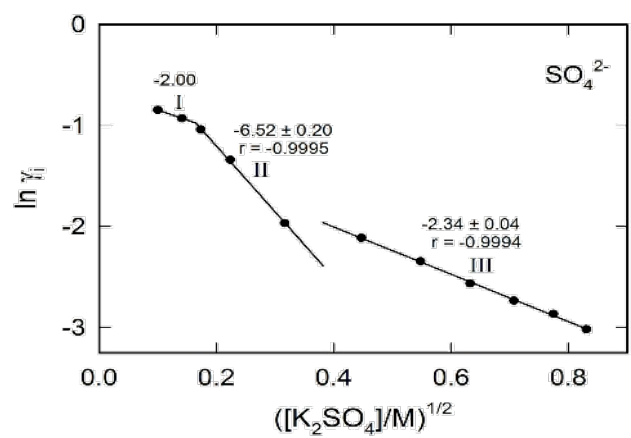

Figs 16A and 16B. Data from Table 5.
$K^+$: Four phases. Transitions: 0.17, 0.45, 0.77-0.83.
$SO_4^{2-}$: Three phases. Transitions: 0.17, 0.32-0.45 (large jump).

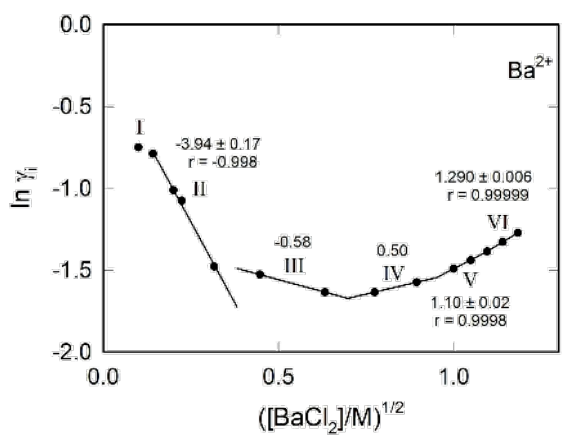 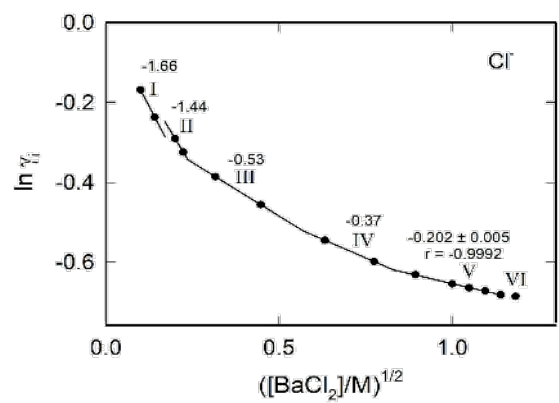

Figs 17A and 17B. Data from Table 6.
$Ba^{2+}$: Six phases. Transitions: 0.10-0.14, 0.32-0.45 (jump), 0.70, 0.95, 1.10. A 3-point line III will have r = -0.994.
$Cl^-$: Six phases. Transitions: 0.14-0.20 (jump), 0.26, 0.57, 0.82, 1.05-1.10 (jump, not visible), 1.14-1.18. Lines V and VI are parallel. Phases VI and VII may be a single phase, but r = -0.975.

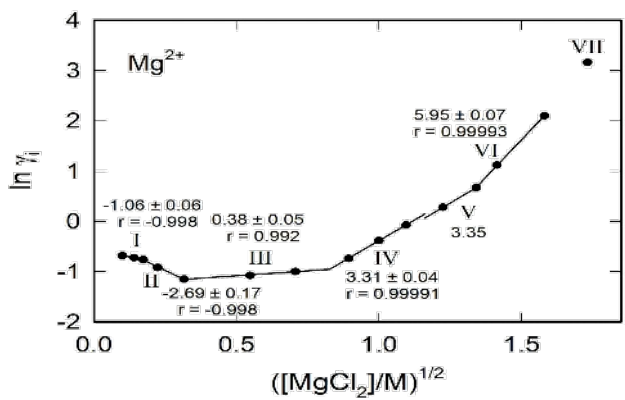 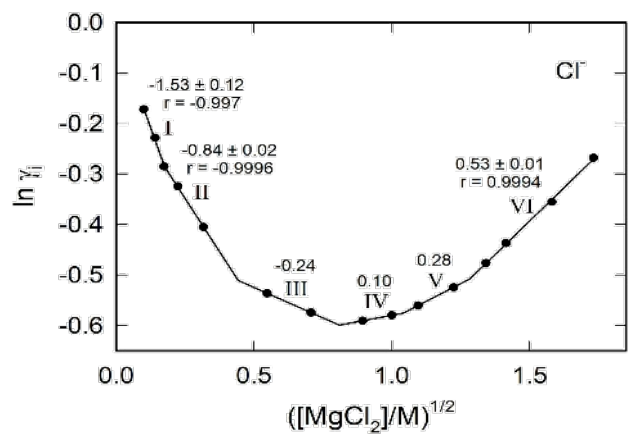

Figs 18A and 18B. Data from Table 6.
$Mg^{2+}$: Seven phases. Transitions: 0.17, 0.33, 0.83, 1.10-1.22 (tiny jump), 1.34, 1.58-1.73. Lines IV and V are parallel.
$Cl^-$: Six phases. Transitions: 0.17, 0.44, 0.81, 1.04, 1.28.

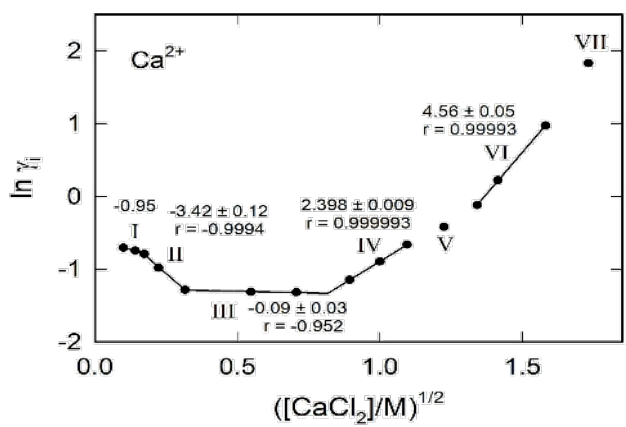 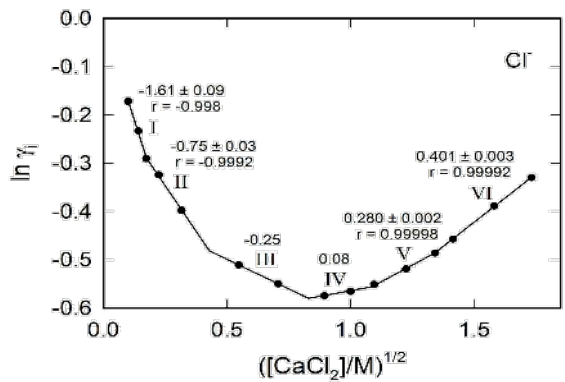

Figs 19A and 19B. Data from Table 6.
$Ca^{2+}$: Seven phases. Transitions: 0.16, 0.32 (just above point 5, r = -0.951 for a horizontal 3-point line III), 0.82, 1.10-1.22 (tiny jump), 1.34, 1.58-1.73. Lines IV and V are about parallel.
$Cl^-$: Six phases. Transitions: 0.17, 0.43, 0.83, 1.09, 1.34.

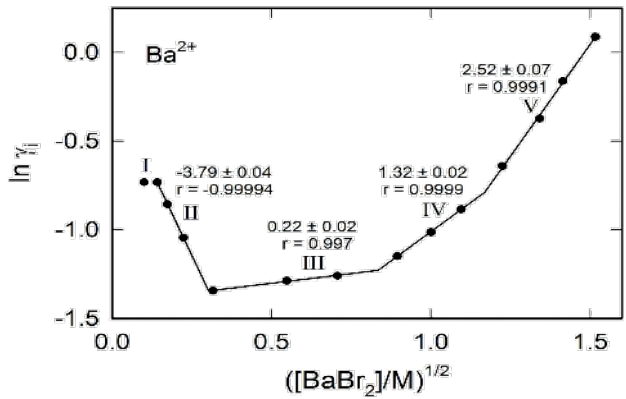
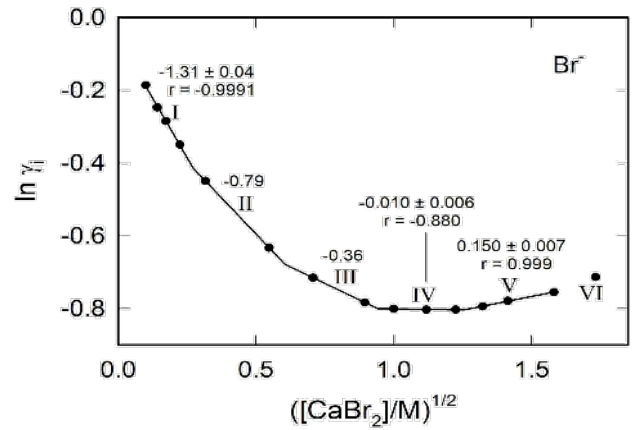

Figs 20A and 20B. Data from Table 7.
$Ba^{2+}$: Five phases. Transitions: 0.10-0.14, 0.30, 0.83, 1.17.
$Br^-$: Seven phases. Transitions: 0.10-0.14, 0.22-0.32 (tiny jump), 0.63, 0.97, 1.22-1.34 (tiny jump), 1.41-1.52. Lines II and III are parallel, as are lines V and VI.

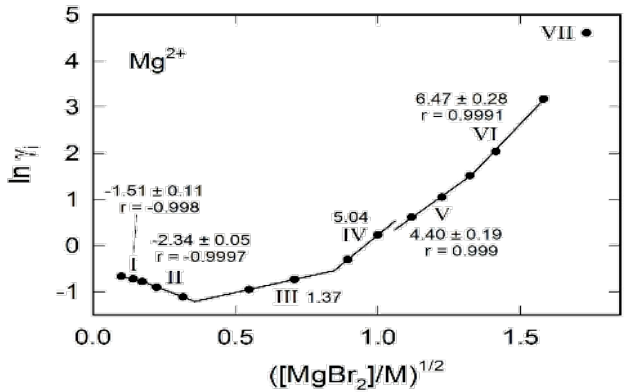
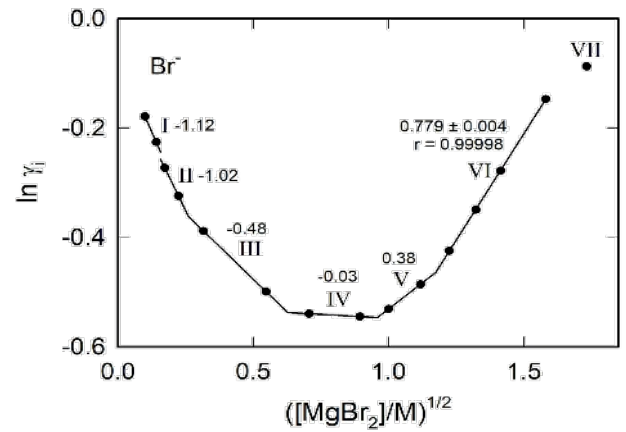

Figs 21A and 21B. Data from Table 7.
$Mg^{2+}$: Seven phases. Transitions: 0.17, 0.36, 0.85, 1.00-1.12 ( jump), 1.32, 1.58-1.73.
$Br^-$: Seven phases. Transitions: 0.14-0.17 (tiny jump), 0.26, 0.63, 0.96, 1.17, 1.58-1.73. Lines I and II are parallel.

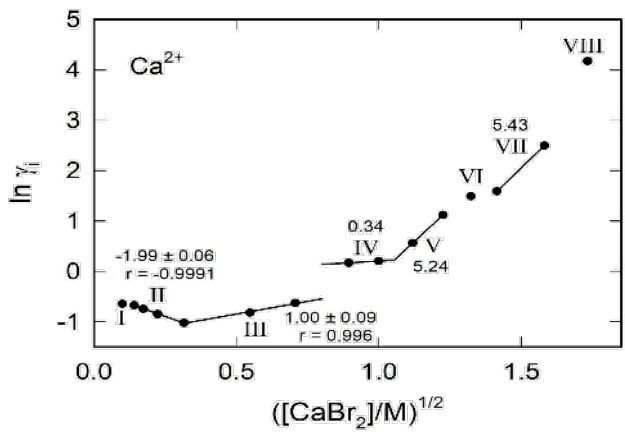
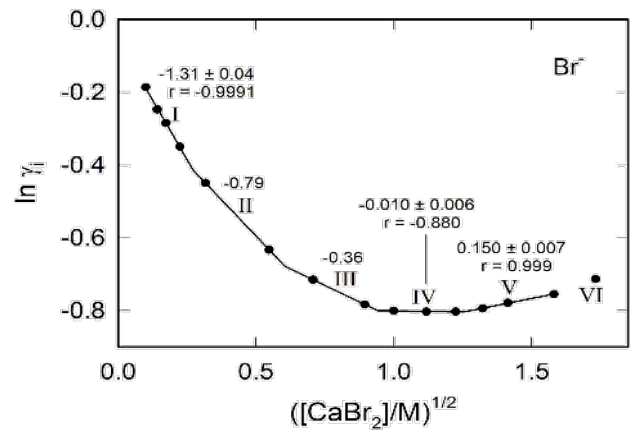

Figs 22A and 22B. Data from Table 7.
$Ca^{2+}$: Eight (or seven) phases. Transitions: 0.10-0.14, 0.32, 0.71-0.89 (large jump), 1.05, 1.22-1.32 (not for seven phases), 1.32-1.41, 1.58-1.73. Lines V and VII are about parallel.
$Br^-$: Six phases. Transitions: 0.27, 0.61, 0.94, 1.25, 1.58-1.73.

# Comparisons

It should first be noted that all the profiles are precisely multiphasic, including the profiles for NaCl (Fig. 4) and CaCl$_2$ (Fig. 19) that have previously (Liu and Eisenberg 2015) been represented as curvilinear.

## KCl, KBr, KF, NaCl, NaBr, NaF

For KCl the plots for cation and anion activities decrease at low to intermediate concentrations befor leveling off or increasing slightly at high concentrations. This may be the case also for KBr and KF, but the data are somewhat limited and there are some differences. At high concentrations, the y values for NaCl and NaBr increase more for the cations than for the anions. The intermediate peak in the cation profile for NaBr is also found in data of Lee et al. (2002) (Nissen 2017b). The data for NaF are limited, and there may be no real difference between the profiles.

The large difference between the cation plots for NaBr and KBr, also noted by Lin and Lee (2003), may help show which effects are most important in theories of activity.

## LiCl, LiBr, CsCl

After an initial decrease for LiCl and LiBr, the y values for cation activities increase much more at high concentrations than do the values for anion activities. For CsCl the values for the two activities are about the same and they decrease consistently with increasing concentrations.

## NaOH, KOH, HCl

For NaOH there is an initial marked decrease in both activities but then a slight increase at high concentrations. For KOH there is a decrease in cation activities, but a marked increase in anion activities. At low concentrations the multiphasic properties of the profile are particularly pronounced. After an initial decline, the increase in cation activities for HCl is greater than the increase in anion activities.

## NaNO$_3$, KNO$_3$, Na$_2$SO$_4$, K$_2$SO$_4$

The NaNO$_3$ data for cation activities may be in error at high concentrations, as indicated by the consistent and marked decrease in the other activities for NaNO$_3$ and KNO$_3$. For the sulfates, the large decreases in anion activities at high concentrations are remarkable.

## BaCl$_2$, MgCl$_2$, CaCl$_2$, BaBr$_2$, MgBr$_2$, CaBr$_2$

There is an initial decrease in cation activities, but a large increase at higher concentrations for MgCl$_2$ and CaCl$_2$. For MgBr$_2$ and CaBr$_2$ there are again very large increases in cation activities at high concentrations. There are also increases in most anion activities, but much smaller.

## Conclusions and Questions

The analysis here, and in preceding papers (Nissen 2017a,b), shows that activation profiles have discontinuous transitions between straight line segments, presumably representing transitions from state to state of the underlying process and perhaps structure. Chemical kinetic models have no shortage of states, particularly models of enzymes (Dixon and Webb 1979) but transitions are traditionally considered smooth (except in the notable case of currents through single ion channels - Neher 1997, Sakmann and Neher 1995). The results reported here are not smooth; they have dramatic discontinuities. The data themselves are remarkable precise, as shown by the correlation functions as reported here (for each of the straight line segments, individually). The discontinuities do not arise from experimental error, in the normal sense of that phrase (as additive instrumentation noise), but from some stochastic property (or extreme sensitivity) of the underlying molecular mechanism itself.

The multiphasic nature of the activation profiles has seemed surprising for many years (Nissen 1971, 1974, 1991, 1996), but it fits beautifully with recent work in structural biology. The multiphasic profiles seem analogous to "the conformation" switching (Ha and Loh 2012) between the intermediary states found in so many protein structures, particularly those observed with cryo-electronmicroscopy (Autzen et al. 2018, Kim and Chen 2018, Saotome et al. 2017, Twomey et al. 2017, Yoder et al. 2018). The measurements of ion activity considered here do not involve proteins. Here, "the conformation" presumably means the number of water molecules packed into the hydration shell(s) of the ion coupled to the surrounding ionic atmosphere. It is possible that the theory of Liu and Eisenberg can be extended to deal with these discrete transitions, since the theory depends sensitively on the number of waters in the inner shell and surrounding annulus, and steric packing effects might make that number a discontinuous function of bulk concentration.

These speculations are just that. The molecular basis of the discontinuities is not known. Why should there be clear and reproducible discontinuities in simple salt solutions? Why should the lines be straight, apparently perfectly so? Why are adjacent lines quite often parallel or nearly so? Why do the profiles for NaBr and KBr differ so markedly? To what extent does the finding of multiphasic profiles invalidate conclusions drawn from averaged curvilinear results?

**Acknowledgment -** We thank Stig Nissen for his help in submitting this and other papers to arXiv.

# References

Autzen HE, Myasnikov AG, Campbell MG, Asarnow D, Julius D, Cheng Y (2018) Structure of the human TRPM4 ion channel in a lipid nanodisc. *Science* 359: 228-232.

Dixon M, Webb EC (1979) *Enzymes*. Academic Press, New York.

Ha J-H, Loh SN (2012) Protein conformational switches: From nature to design. *Chemistry* 18: 7984-7999.

Kim Y, Chen J (2018) Molecular structure of human P-glycoprotein in the ATP-bound, outward-facing conformation. *Science* 359: 915-919.

Lee L-S, Chen T-M, Tsai K-M, Lin C-L (2002) Individual ion activity and mean activity coefficients in NaCl, NaBr, KCl and KBr aqueous solutions. *J Chin Inst Chem Engrs* 33: 267-281.

Lin C-L, Lee L-S (2003) A two-ionic-parameter approach for ion activity coefficients of aqueous electrolyte solutions. *Fluid Phase Equilibria* 205: 69-88.

Liu J-L, Eisenberg B (2015) Poisson-Fermi model of single ion activities in aqueous solutions. *Chem Phys Lett* 637: 1-6.

Neher E (1997) *Nobel Lectures, Physiology or Medicine 1991-1995*. (Ringertz N, ed.). World Scientific Publishing Co, Singapore, pp. 10.

Nissen P (1971) Uptake of sulfate by roots and leaf slices of barley. *Physiol Plant* 24: 315-324.

Nissen P (1974) Uptake mechanisms: Inorganic and organic. *Annu Rev Plant Physiol* 25: 53-79.

Nissen P (1991) Multiphasic uptake mechanisms in plants. *Int Rev Cytol* 126: 89-134.

Nissen P (1996) Uptake mechanism. Pp. 511-527 in: Waisel Y, Eshel A, Kafkafi U (eds). Plant Roots. The Hidden Half (2. Ed). Marcel Dekker, Inc, New York.

Nissen P (2015a) Discontinuous transitions: Multiphasic profiles for channels, binding, pH, folding and chain length. Posted on arXiv.org with Paper ID arXiv:1511.06601.

Nissen P (2015b) Multiphasic pH profiles for the reaction of tris-(hydroxymethyl)-aminomethane with phenyl esters. Posted on arXiv.org with Paper ID arXiv:1512.02561.

Nissen P (2016a) Profiles for voltage-activated currents are multiphasic, not curvilinear. Posted on arXiv.org with Paper ID arXiv:1603.05144.

Nissen P (2016b) Multiphasic profiles for voltage-dependent $K^+$ channels: Reanalysis of data of MacKinnon and coworkers. Posted on arXiv.org with Paper ID arXiv: 1606.02977.

Nissen P (2016c) 'Perfectly' curvilinear profiles for binding as determined by ITC may in fact be multiphasic. Posted on arXiv.org with Paper ID arXiv: 1606.09133.

Nissen P (2016d) Multiphasic interactions between nucleotides and target proteins. Posted on arXiv.org with Paper ID arXiv: 1608.07459.

Nissen P (2017a) Multiphasic profiles for the biologically important ion activities of NaCl and KCl. Posted on arXiv.org with paper ID arXiv: 1712.01758.

Nissen P (2017b) Multiphasic profiles for the ion activities of NaBr and KBr. Posted on arXiv.org with paper ID arXiv: 1712.07012.

Nissen P, Eisenberg B (2018) Profiles for ion activities are multiphasic, not curvilinear. ArXiv. To be submitted.

Nissen P, Martín-Nieto J (1998) 'Multimodal` kinetics: Cyanobacteria nitrate reductase and other enzyme, transport and binding systems. *Physiol Plant* 104: 503-511.

Sakmann B, Neher E (1995) *Single Channel Recording*. Plenum, New York. Second edn.

Saotome K, Murthy SE, Kefauver JM, Whitwam T, Patapoutian A, Ward AB (2017) Structure of the mechanically activated ion channel Piezo1. *Nature* 554: 481-486.

Twomey EC, Yelshanskaya MV, Grassucci RA, Frank J, Sobolevsky AI (2017) Channel opening and gating mechanism in AMPA-subtype glutamate receptors. *Nature* 549: 60-65.

Wilczek-Vera G, Rodil E, Vera JH (2004) On the activity of ions and the junction potential: Revised values for all data. *AIChe Journal* 50: 445-462.

Yoder N, Yoshioka C, Gouaux E (2018) Gating mechanism of acid-sensing ion channels. *Nature* 555: 397-401.